

\def\singlespace{\normalbaselines}
\def\oneandahalfspace{\baselineskip=1.15\normalbaselineskip plus 1pt
\lineskip=2pt\lineskiplimit=1pt}

\def\np{\vfill\eject}

\def\nofirstpagenoten{\nopagenumbers\footline={\ifnum\pageno>1\tenrm
\hss\folio\hss\fi}}
\def\nofirstpagenotwelve{\nopagenumbers\footline={\ifnum\pageno>1\twelverm
\hss\folio\hss\fi}}
\def\leaderfill{\leaders\hbox to 1em{\hss.\hss}\hfill}
\def\ft#1#2{{\textstyle{{#1}\over{#2}}}}
\def\frac#1/#2{\leavevmode\kern.1em
\raise.5ex\hbox{\the\scriptfont0 #1}\kern-.1em/\kern-.15em
\lower.25ex\hbox{\the\scriptfont0 #2}}
\def\sfrac#1/#2{\leavevmode\kern.1em
\raise.5ex\hbox{\the\scriptscriptfont0 #1}\kern-.1em/\kern-.15em
\lower.25ex\hbox{\the\scriptscriptfont0 #2}}


\parindent=20pt
\def\narrow{\advance\leftskip by 40pt \advance\rightskip by 40pt}

\def\AB{\bigskip
        \centerline{\bf ABSTRACT}\medskip\narrow}
\def\nonarrower{\advance\leftskip by -40pt\advance\rightskip by -40pt}
\def\AE{\bigskip\nonarrower}

\def\boxit#1{\vbox{\hrule\hbox{\vrule\kern3pt
        \vbox{\kern3pt#1\kern3pt}\kern3pt\vrule}\hrule}}

\def\gtorder{\mathrel{\raise.3ex\hbox{$>$}\mkern-14mu
             \lower0.6ex\hbox{$\sim$}}}
\def\ltorder{\mathrel{\raise.3ex\hbox{$<$}|mkern-14mu
             \lower0.6ex\hbox{\sim$}}}
\def\dalemb#1#2{{\vbox{\hrule height .#2pt
        \hbox{\vrule width.#2pt height#1pt \kern#1pt
                \vrule width.#2pt}
        \hrule height.#2pt}}}

\font\fourteentt=cmtt10 scaled \magstep2
\font\fourteenbf=cmbx12 scaled \magstep1
\font\fourteenrm=cmr12 scaled \magstep1
\font\fourteeni=cmmi12 scaled \magstep1
\font\fourteenss=cmss12 scaled \magstep1
\font\fourteensy=cmsy10 scaled \magstep2
\font\fourteensl=cmsl12 scaled \magstep1
\font\fourteenex=cmex10 scaled \magstep2
\font\fourteenit=cmti12 scaled \magstep1
\font\twelvett=cmtt10 scaled \magstep1 \font\twelvebf=cmbx12
\font\twelverm=cmr12 \font\twelvei=cmmi12
\font\twelvess=cmss12 \font\twelvesy=cmsy10 scaled \magstep1
\font\twelvesl=cmsl12 \font\twelveex=cmex10 scaled \magstep1
\font\twelveit=cmti12
\font\tenss=cmss10
 
 \font\ninebf=cmbx7 scaled \magstep1
\font\ninerm=cmr7 scaled \magstep1 \font\ninei=cmmi7 scaled \magstep1
\font\ninesy=cmsy7 scaled \magstep1 
\font\eightrm=cmr7 scaled 1140 
 
\font\sevenbf=cmbx7 \font\sevenrm=cmr7 \font\seveni=cmmi7
\font\sevensy=cmsy7 

\catcode`@=11
\newskip\ttglue
\newfam\ssfam

\def\fourteenpoint{\def\rm{\fam0\fourteenrm}
\textfont0=\fourteenrm \scriptfont0=\tenrm \scriptscriptfont0=\sevenrm
\textfont1=\fourteeni \scriptfont1=\teni \scriptscriptfont1=\seveni
\textfont2=\fourteensy \scriptfont2=\tensy \scriptscriptfont2=\sevensy
\textfont3=\fourteenex \scriptfont3=\fourteenex \scriptscriptfont3=\fourteenex
\def\it{\fam\itfam\fourteenit} \textfont\itfam=\fourteenit
\def\sl{\fam\slfam\fourteensl} \textfont\slfam=\fourteensl
\def\bf{\fam\bffam\fourteenbf} \textfont\bffam=\fourteenbf
\scriptfont\bffam=\tenbf \scriptscriptfont\bffam=\sevenbf
\def\tt{\fam\ttfam\fourteentt} \textfont\ttfam=\fourteentt
\def\ss{\fam\ssfam\fourteenss} \textfont\ssfam=\fourteenss
\tt \ttglue=.5em plus .25em minus .15em
\normalbaselineskip=16pt
\abovedisplayskip=16pt plus 4pt minus 12pt
\belowdisplayskip=16pt plus 4pt minus 12pt
\abovedisplayshortskip=0pt plus 4pt
\belowdisplayshortskip=9pt plus 4pt minus 6pt
\parskip=5pt plus 1.5pt
\setbox\strutbox=\hbox{\vrule height12pt depth5pt width0pt}
\let\sc=\tenrm
\let\big=\fourteenbig \normalbaselines\rm}
\def\fourteenbig#1{{\hbox{$\left#1\vbox to12pt{}\right.\n@space$}}}

\def\twelvepoint{\def\rm{\fam0\twelverm}
\textfont0=\twelverm \scriptfont0=\ninerm \scriptscriptfont0=\sevenrm
\textfont1=\twelvei \scriptfont1=\ninei \scriptscriptfont1=\seveni
\textfont2=\twelvesy \scriptfont2=\ninesy \scriptscriptfont2=\sevensy
\textfont3=\twelveex \scriptfont3=\twelveex \scriptscriptfont3=\twelveex
\def\it{\fam\itfam\twelveit} \textfont\itfam=\twelveit
\def\sl{\fam\slfam\twelvesl} \textfont\slfam=\twelvesl
\def\bf{\fam\bffam\twelvebf} \textfont\bffam=\twelvebf
\scriptfont\bffam=\ninebf \scriptscriptfont\bffam=\sevenbf
\def\tt{\fam\ttfam\twelvett} \textfont\ttfam=\twelvett
\def\ss{\fam\ssfam\twelvess} \textfont\ssfam=\twelvess
\tt \ttglue=.5em plus .25em minus .15em
\normalbaselineskip=14pt
\abovedisplayskip=14pt plus 3pt minus 10pt
\belowdisplayskip=14pt plus 3pt minus 10pt
\abovedisplayshortskip=0pt plus 3pt
\belowdisplayshortskip=8pt plus 3pt minus 5pt
\parskip=3pt plus 1.5pt
\setbox\strutbox=\hbox{\vrule height10pt depth4pt width0pt}
\let\sc=\ninerm
\let\big=\twelvebig \normalbaselines\rm}
\def\twelvebig#1{{\hbox{$\left#1\vbox to10pt{}\right.\n@space$}}}

\def\tenpoint{\def\rm{\fam0\tenrm}
\textfont0=\tenrm \scriptfont0=\sevenrm \scriptscriptfont0=\fiverm
\textfont1=\teni \scriptfont1=\seveni \scriptscriptfont1=\fivei
\textfont2=\tensy \scriptfont2=\sevensy \scriptscriptfont2=\fivesy
\textfont3=\tenex \scriptfont3=\tenex \scriptscriptfont3=\tenex
\def\it{\fam\itfam\tenit} \textfont\itfam=\tenit
\def\sl{\fam\slfam\tensl} \textfont\slfam=\tensl
\def\bf{\fam\bffam\tenbf} \textfont\bffam=\tenbf
\scriptfont\bffam=\sevenbf \scriptscriptfont\bffam=\fivebf
\def\tt{\fam\ttfam\tentt} \textfont\ttfam=\tentt
\def\ss{\fam\ssfam\tenss} \textfont\ssfam=\tenss
\tt \ttglue=.5em plus .25em minus .15em
\normalbaselineskip=12pt
\abovedisplayskip=12pt plus 3pt minus 9pt
\belowdisplayskip=12pt plus 3pt minus 9pt
\abovedisplayshortskip=0pt plus 3pt
\belowdisplayshortskip=7pt plus 3pt minus 4pt
\parskip=0.0pt plus 1.0pt
\setbox\strutbox=\hbox{\vrule height8.5pt depth3.5pt width0pt}
\let\sc=\eightrm
\let\big=\tenbig \normalbaselines\rm}
\def\tenbig#1{{\hbox{$\left#1\vbox to8.5pt{}\right.\n@space$}}}
\let\rawfootnote=\footnote \def\footnote#1#2{{\rm\parskip=0pt\rawfootnote{#1}
{#2\hfill\vrule height 0pt depth 6pt width 0pt}}}

\def\tenfoot{\tenpoint\hskip-\parindent\hskip-.1cm}

\overfullrule=0pt
\twelvepoint
\oneandahalfspace
\def\sbullet{\raise.2em\hbox{$\scriptscriptstyle\bullet$}}
\nofirstpagenotwelve
\baselineskip 15pt

\def\ft#1#2{{\textstyle{{#1}\over{#2}}}}

\def\ul{\underline}


\rightline{CTP TAMU--28/93}
\rightline{hep-th/9310126}
\rightline{October 1993}
\vskip 2truecm
\centerline{\bf Aspects of $\kappa$--symmetry\footnote{$^*$}{\tenfoot \sl
Based on a talk presented at the {\it Salamfest}, ICTP, Trieste,
March, 1993.}}
\vskip 1.5truecm
\centerline{E. Sezgin\footnote{$^\dagger$}{\tenfoot
\sl  Supported in part by the National Science Foundation, under grant
PHY-9106593.}}

\vskip 1.5truecm
\centerline{\it  Center for Theoretical Physics, Texas A\&M University}
\centerline{\it College Station, TX 77843--4242, USA }
\vskip 1.5truecm


\AB\singlespace
 {\tenfoot We review various aspects of a fermionic gauge symmetry, known as
the
$\kappa$--symmetry, which plays an important role in formulations of
superstrings, supermembranes and higher dimensional extended objects.
We also review some aspects of the connection between $\kappa$--symmetric
theories
and their twistor-like formulations. }
\AE
\np
\oneandahalfspace
\noindent{\bf 1. Introduction}
\bigskip

$\kappa$--symmetry [1] is a remarkable symmetry which plays an important role
in
the manifestly spacetime supersymmetric formulation of string theories [2].
It is also crucial for the existence of super $p$--brane theories [3][4][5].
While
for $p=1$ (string) there exists a formulation which does not have manifest
spacetime supersymmetry but has a world-sheet local supersymmetry (the NSR
formulation), such a formulation is forbiddingly complicated for higher super
$p$--branes.

$\kappa$--symmetry is a fermionic symmetry, and in most models where it
occurs, interestingly enough, a gauge field is not necessary for its
realization.
Furthermore, $\kappa$--symmetry is typically an on-shell symmetry and
consequently, there is no simple way to construct higher order
$\kappa$--invariants. Whether $\kappa$--symmetry is powerful enough to
improve dramatically  the finiteness behaviour of higher super $p$--branes is
probably one of the most interesting and important open questions.
A formidable obstacle to answering this question is the absence of a well
defined
covariant quantization scheme for $\kappa$--symmetry and this is basically
due to
the fact that the BRST quantization of the symmetry requires an infinite
number of
ghosts [6].

	The aim of this relatively short review is to outline some of the
salient
features of $\kappa$--symmetric field theories, and some of the outstanding
open problems. This article is based on a talk given at the conference in
honor of Professor Abdus Salam. It is a great pleasure to make this
contribution. Professor Abdus Salam has made so many important contributions
in so
many areas that I did not have any difficulty in choosing a
subject which owes greatly to his work.  As we will see below,
$\kappa$--symmetric field theories are essentially sigma models formulated in a
superspace, and the background fields occurring in these models are described
by
superfields. While these models have a very simple and elegant geometric form
in superspace [7][8], it  would be  a nightmare to describe them in component
formalism, in presence of a curved background. The notions of superspace and
superfields, which play such a significant role in these theories, were
invented by
Salam and Strathdee in 1974 [9].  The concluding remark in their paper is
rather
interesting and  it reflects their admirable approach to doing physics and
their
humility even when they present an important result [9]:

{\it ``The approach discussed in this paper {\rm [the superspace approach]}
may not
provide the most serviceable one available but, with the present hazy
understanding
of this curious and potentially important symmetry {\rm [supersymmetry]},
it seems
worthwhile to examine every avenue''}.

\bigskip
\noindent{\bf 2. Massless Superparticle in Curved Superspace}
\bigskip
As mentioned above, $\kappa$--symmetric actions are basically sigma models
in which
the target  space is a superspace. We take it to be a Salam-Strathdee type
superspace which can be viewed as the supercoset  $G/H$ where $G$ is the super
Poincar\'e group and $H$ is the Lorentz group in $d$ dimensions [9]. It would
be very interesting to consider more general supermanifolds. Focusing on
the usual scenario, we consider a target superspace with coordinates
$Z^M=(X^m,\theta^\mu)$. It is useful to define the pull-back supervielbein as
$$
E_\tau^A= \partial_\tau Z^M E_M{}^A\ , \eqno(2.1)
$$
where $\partial_\tau$ denotes differentiation with respect to the worldline
time
variable. The tangent space index splits as $A=(a,\alpha)$, where
$a=0,1,...,d-1$ labels a Lorentz vector and $\alpha$ labels the fundamental
representation of the Lorentz group $\times$ the automorphism group of the
super
Poincar\'e algebra in $d$ dimensions. The simplest action for massless
superparticle is given by
$$
S= \ft12\int d\tau e^{-1}\phi E_\tau^aE_\tau^a\ , \eqno(2.2)
$$
where $e$ is the einbein field, $\phi(Z)$ is the dilaton superfield and
it is  understood that the contraction of the
Lorentz indices is with the Lorentz metric $\eta_{ab}$. Note that this action
describes the coupling of a massless particle to the target space fields that
are contained in the superfields $E_M{}^A$ and $\phi$. (In flat target space,
this action reduces to the Brink-Schwarz superparticle action [10]). The
remarkable
property of this action is that, with suitable constraints imposed on the
target
superspace torsion, it is invariant under the so-called $\kappa$--symmetry
transformations which take the form [1][8]
$$
\delta Z^M=\kappa_\alpha E_\tau^a\Gamma_a^{\alpha\beta}E_\beta{}^M\ ,\quad\quad
e^{-1}\delta e =  \kappa_\alpha S^\alpha\ , \eqno(2.3)
$$
where $S^\alpha$ is a function of the superfields to be determined by the
$\kappa$--symmetry of the action and $E_A{}^M$ is the inverse supervielbein
satisfying $E_A{}^M E_M{}^B=\delta_A^B$. The torsion tensor is defined by
$T^A=dE^A+E^B\wedge \omega_B{}^A=\ft12 E^B\wedge E^C T_{CB}{}^A$ where the
basis
one form is $E^A=dZ^M E_M{}^A$ and the spin connection one form is
$\omega_A^{~B}=
dZ^M\omega_{MA}{}^B$. In standard superspace the tangent space group is
Lorentzian
so that $\omega_a{}^\alpha=0$ and
$\omega_\alpha{}^a=0$. Keeping this in mind, and making use of the following
lemma
$$
\delta E_\tau^A=\partial_\tau\big(\delta Z^M E_M{}^A \big)
+\delta Z^M E_M{}^B E_\tau^C\bigl(
-T_{CB}{}^A+\omega_C{}^A{}_B - (-1)^{BC}\omega_B{}^A{}_C \big)\ ,\eqno(2.4)
$$
we find that the invariance of the action (2.2) under the $\kappa$--symmetry
transformations (2.3) requires the following torsion constraints
$$
T_{\alpha\beta}{}^c =2(\Gamma^c)_{\alpha\beta}\ ,\quad\quad\quad
T_{\alpha (bc)} =u^\beta{}_{(b}\Gamma_{c)\beta\alpha}+\eta_{bc}
v_\alpha\ ,\eqno(2.5)
$$
and fixes the function $S^\alpha$ to be
$$
S^\alpha=-4E_\tau^\alpha+E_\tau^a\big(2 u^\alpha{}_a
+2\Gamma_a^{\alpha\beta}v_\beta +\Gamma_a^{\alpha\beta}\phi^{-1}D_\beta
\phi\big)\ ,
\eqno(2.6)
$$
where $u^\alpha{}_{a}$ is an arbitrary  {\it $\Gamma$--traceless} vector-spinor
superfield
\footnote{$^\dagger$} {\tenfoot The freedom for having a vector-spinor
in the $\kappa$--symmetry imposed constraint on $T_{\alpha(ab)}$ is usually
overlooked, but  as one allows field redefinitions afterwards to arrive  at a
standard set of constraints, this omission becomes inconsequential.}
and $v^\alpha$ is an arbitrary spinor superfield.

   The action is also invariant under the worldline reparametrizations,
and provided that $u^\beta{}_a=0$ and $\phi v_\alpha+\ft12 D_\alpha\phi=0 $,
also under the local  $\lambda$--symmetry transformations
$$
\delta Z^M= \lambda E_\tau^\alpha E_\alpha{}^M\ ,\quad\quad\quad
\delta e=0\ , \eqno(2.7)
$$
where $\eta(\tau)$ and $\lambda(\tau)$ are arbitrary parameters. (For flat
target
space version of (2.7), see ref. [2]).  Furthermore, provided that the target
superspace admits a {\it superconformal  Killing vector $k^M$}, the action
is also
invariant under the following  target space rigid superconformal
transformations
[11]
$$
\delta Z^M= k^M\ ,\qquad\qquad \delta e= 2e(U+\ft12 k^M\partial_M \phi)\ ,
\eqno(2.8)
$$
where $U$ is a scalar superfield which occurs in the
superconformal Killing equation ${\cal L}_k E_M{}^a=UE_M{}^a+E_M{}^b L_b{}^a$,
where the second term is a Lorentz transformation. Superconformal Killing
vectors
can be found at least in certain spacetimes of dimension  $d\le 7$. In higher
dimensions one can always find the super Killing vectors which obey the super
Poincar\'e algebra. The action will be invariant under such
transformations, and hence its manifest spacetime supersymmetry.

	The algebra of the symmetry transformations described above closes
on-shell,
i.e. modulo the equations of motion which take the form
$$
\eqalign{
&E_\tau^aE_\tau^a=0\ , \cr
&E_\tau^a\Gamma^a_{\alpha\beta}S^\beta=0 \ , \cr
&e\phi^{-1}\partial_\tau\bigl
(e^{-1}\phi E_\tau^a\bigr) +E_\tau^bE_\tau{}^c\bigl( -T_{abc}+\omega_{bca}\bigr
) +E_\tau^bE_\tau^\alpha\bigl (T_{\alpha ab}-\omega_{\alpha ab}\bigr )=0\ .\cr}
\eqno(2.9)
$$

Instead of deriving the algebra of the $\kappa$--transformations
directly in the Lagrangian formalism, following [12],  we shall consider the
algebra of first class constraints which generate the
$\kappa$--transformations. To
this end let us define the conjugate momenta
$$
\eqalign{
P_A  &=E_A{}^M{\delta S\over \delta \partial_\tau Z^M}\ , \cr
      &=e^{-1}\phi E_\tau^a\delta_A^a\ .  \cr}\eqno(2.10)
$$
The constraints which follow from the form of the Lagrangian are
$P_aP^a\approx 0$
and $P_\alpha\approx 0$. However, these do not form a set of first class
constraints and therefore do not form a closed algebra. Instead, let us
define the
combinations [12][13]

$$
{\cal A} =\ft12 P^a P_a +A^\alpha P_\alpha\ , \quad\quad\quad
           {\cal B}^\alpha = A^{\alpha\beta}P_\beta\ ,  \eqno(2.11)
$$
where $A^{\alpha}$ and $A^{\alpha\beta}$ are functions of superfields which
are to
be determined by the closure of the algebra of the above constraints. For
definiteness let us consider the {\it ten dimensional spacetime}. It turns
out that
$A^\alpha=0$ and $A^{\alpha\beta}=\Gamma_a^{\alpha\beta}P_a$. Furthermore,
in [12]
it has been shown that various sets of constraints can be obtained from those
required by the closure of the constraint algebra plus the Bianchi identity
$\sum_{(ABC)} \bigl( D_A T_{BC}{}^D - R_{ABC}{}^D +
T_{AB}{}^ET_{EC}{}^D \bigr)=0$, where $R_{ABC}{}^D$ is the Riemann curvature.
In
addition, if one  allows the local Lorentz transformations  $\delta
E^a=E^b\Lambda_b{}^a$ and transformations of the form  $\delta
E^\alpha=E^b\Lambda_b{}^\alpha+E^\beta \Lambda_\beta{}^\alpha$, it has been
shown
that [12] the set of constraints proposed by Nilsson [14]
$$
\eqalignno{
&T_{\alpha\beta}{}^a=2(\Gamma^a)_{\alpha\beta}\ , \quad\quad   T_{\alpha a}{}^b
=v_\alpha\delta_a^b\ ,\quad\quad T_{ab}{}^c=0\ , \cr
&T_{\alpha\beta}{}^\gamma=(\Gamma^{abcde})_{\alpha\beta}
             (\Gamma_{abcde})^{\gamma\delta}\ v_\delta\ ,  &(2.12)\cr}
$$
or the set proposed by Witten [8]
$$
\eqalignno{
&T_{\alpha\beta}{}^a =
2(\Gamma^a)_{\alpha\beta}\ , \quad\quad  T_{\alpha a}{}^b=0\ , \quad\quad
                       T_{a(bc)}=0\ , \cr
&T_{\alpha\beta}{}^\gamma=0\ ,
\quad\quad T_{a\alpha}{}^\beta=
-\ft1{24}\bigl(\Gamma_a\Gamma^{bcd}\bigr)_\alpha{}^\beta T_{bcd}\ ,
&(2.13)\cr}
$$
can be derived.  In obtaining (2.13) scale transformations
of the form $\delta E^\alpha=-\ft12 u E^\alpha$ and $\delta E^a=-u E^a$
must also
be allowed.

The two sets of constraints (2.12) and (2.13), as well as many other possible
choices of constraints are all equivalent via appropriate field redefinitions.
It is interesting to see how one such particular set, which differs from (2.12)
and (2.13), emerges from the principle of light-like integrability in loop
superspace [15]. In fact, one expects a very close connection between
light-like
integrability and $\kappa$--symmetry [8][12][15].

Neither set of the
constraints (2.12) and (2.13) are  sufficient to imply the $N=1, d=10$
supergravity
equations of motion.  To describe them, one has to introduce a 3-form
superfield in Nilsson's case, while in Witten's case, as $T_{abc}$ happens to
be
totally antisymmetric, one can construct a super 3-form $H$ in terms of
$T_{abc}$
and the dilaton superfield $\phi$ [8][16]. Either set of constraints turn out
to
arise in superstring theory where the string couples to the two form field $B$
whose  field strength is $H=dB$.

	The massless superparticle can also be coupled to background Maxwell
and Yang-Mills
fields. For simplicity let us take the gauge group to be $SO(N)$. Introducing
fermionic worldline fields $\psi^I(\tau), I=1,...,N$, the action can be
written as
$$
S= \int d\tau \bigl (\ft12 e^{-1}\phi E_\tau^aE_\tau^a+ \partial_\tau Z^M B_M
+\psi^I D_\tau \psi^I\bigr)\ , \eqno(2.14)
$$
where $D_i\psi^I=\partial_\tau \psi^I+\partial_\tau Z^M A_M^{IJ}\psi^J$\ and
$B_M(Z),\ A_M^{IJ}(Z)$ are the background Maxwell and Yang-Mills
superfields. The $\kappa$--symmetry transformations  again take the
form (2.3) with the additional transformation rule
$$
   \delta \psi^I=-\delta Z^M A_M^{IJ} \psi^J\ .  \eqno(2.15)
$$
The $\kappa$--symmetry transformations leave the action invariant provided that
in addition to the constraints (2.5), the following ones are imposed
$$
H_{\alpha\beta}=0\ , \qquad H_{a\alpha}=\phi(\Gamma_a)_{\alpha\beta}
h^\beta\ ,
\qquad F_{\alpha\beta}^{IJ}=0\ ,\qquad
F_{a\alpha}^{IJ}=(\Gamma_a)_{\alpha\beta}
\chi^{\beta IJ}\ , \eqno(2.16)
$$
and that the function $S^\alpha$ is given by
$$
S^\alpha=-4E_\tau^\alpha+E_\tau^a\big(2 u^\alpha{}_a
+2\Gamma_a^{\alpha\beta}v_\beta+\Gamma_a^{\alpha\beta}\phi^{-1}D_\beta\phi\big)
+\big( h^\alpha+\phi^{-1}\chi^\alpha_{IJ}\psi^I\psi^J\big) \ , \eqno(2.17)
$$
where $h^\alpha,\ \chi^\alpha_{IJ}$ are arbitrary
superfields and $H=dB,\ F=dA+A\wedge A$.  It should be noted that while the
$B$--term in (2.14) is not necessary for $\kappa$--symmetry of the action, it
will become necessary in the case of a {\it massive} superparticle action, as
we
will see in Sec. 4.

The action (2.14) is
also invariant under the worldline reparametrizations, and provided that
$A_M^{IJ}$ and
$B_M$ are symmetric tensors with respect to the superconformal Killing vectors,
also invariant under the target space rigid superconformal symmetry (2.8).  The
$\lambda$--symmetry (2.7) and (2.15), however, imposes the constraints
$h^\alpha=0$ and $\chi^\alpha_{IJ}=0$, in addition to the previous ones:
$u^\alpha{}_a=0,\ \phi v^\alpha+\ft12 D_\alpha \phi =0$.

Considering
again the ten dimensional spacetime, the constraints (2.16) describe super
Maxwell
and super Yang-Mills equations of motion. The coupled supergravity plus  super
Maxwell/super Yang-Mills system does not arise from the constraints implied by
the $\kappa$--symmetry of the superparticle. Superstring theory however does
describe such a system as we shall see later.  In passing we note that
coupling of
Yang-Mills fields to the massless superparticle can be achieved also by using
bosonic coordinates [12] instead of the fermionic ones used above.
 \bigskip
\centerline {\it Flat Target Superspace}
\bigskip
It is useful to consider the flat
target superspace limit of the actions described above. For simplicity let us
consider the action (2.2) with the dilaton field set equal to a constant, e.g.
$\phi=1$.  In a Salam-Strathdee  superspace the spin connection is
vanishing but the torsion does not completely vanish. Its only nonvanishing
component is $T_{\alpha\beta}{}^a=2(\Gamma^a)_{\alpha\beta}$.  Thus from the
definition of the torsion one finds the components of the supervielbein in flat
superspace to be $E_m{}^a=\delta_m^a,\
E_m{}^\alpha=0,\ E_\mu{}^a=(\Gamma^a\theta)_\mu,\
E_\mu{}^\alpha=\delta_\mu^\alpha$. Substituting this into (2.2) gives
$$
S=\ft12\int d\tau e^{-1} \bigl(\partial_\tau
X^a-{\bar\theta}\Gamma^a\partial_\tau\theta\bigr) \bigl(\partial_\tau
X^b-{\bar\theta}\Gamma^b\partial_\tau\theta\bigr)\eta_{ab}\ . \eqno(2.18)
$$
This is the Brink-Schwarz superparticle action [10]. The
$\kappa$--transformation rules (2.3) become [1]
$$
\delta X^a ={\bar\theta}\Gamma^a\delta\theta\ ,\quad\quad \quad
\delta \theta =\Pi^a\Gamma_a \kappa\ ,\quad\quad\quad
e^{-1}\delta e = -4{\bar \kappa}\partial_\tau \theta\ .  \eqno(2.19)
$$
where
$\Pi^a = \bigl(\partial_\tau
X^a-{\bar\theta}\Gamma^a\partial_\tau\theta\bigr)$.
Since on-shell $\Pi^a\Pi_a=0$ and that $ Tr\ \Pi^a
\Gamma_a =0$, the matrix $\Pi^a\Gamma_a=0$ has half as many zero
eigenvalues, and therefore using $\kappa$--symmetry one can gauge away only
half as
many degrees of freedom $\theta^\alpha$.

This phenomenon of halving the fermionic degrees of freedom is rather similar
to
a phenomenon discovered long ago by Mack and Salam [17] in their study of
conformally invariant field theories formulated on a five dimensional cone.
{}From
manifestly $SO(4,2)$ invariant field theories on the cone, by
suitably restricting the fields, and in the case of fermions by a gauge
symmetry
similar to $\kappa$--symmetry, they could obtain the  $3+1$ dimensional
description of correct degrees of freedom. For example, let us define the cone
by $y^A y_A=0,\ A=1,...,6$, where $y^A$ are the embedding coordinates of a
$4+2$
dimensional plane. Consider the manifestly $SO(4,2)$ invariant field equation
for a fermionic field $\psi$ to be of the form:
$$
\bigl(\Gamma^{AB}y_A\partial_B-2\bigr)\psi=0\ , \quad\quad\quad
y^A\partial_A\psi=-2\psi\ ,  \eqno(2.20)
$$
where the second equation is the restriction imposed on the field, representing
its homogeneity degree on the cone. The above field equation is actually
invariant
under the following transformations
$$
\delta \psi= y^A\Gamma_A\kappa\ ,\quad\quad\quad
           y^A\partial_A\kappa=-3\kappa\ ,\eqno(2.21)
$$
where, the second equation expresses the condition which has to be imposed
on the
symmetry parameter $\kappa$ in the form of homogeneity degree on the cone [18].
Since $y^Ay_A=0$\ and\ $Tr\ y^A\Gamma_A =0$, indeed the field $\psi$ has has
as  many
degrees of freedom in $3+1$ dimension. This phenomenon and the form of the
transformation rules is similar to the Siegel's $\kappa$--symmetry
transformation
rules [1]. Given that superparticle actions in $d=3,4,6$ dimensions have
also superconformal symmetry, one wonders if
there is a deeper connection between the two phenomenon and
possibly with the remarkable representations of the superconformal groups
known as
the supersingletons [18].
\bigskip
\centerline{\it Quantization}
\bigskip
In covariant BRST quantization of the massless superparticle, or indeed
any $\kappa$--symmetric system, two features arise. Firstly,
the $\kappa$--transformation rules close on-shell, and second, the
$\kappa$--symmetry is an infinitely reducible type symmetry, in the terminology
of Batalin and Vilkovisky [19]. Both of these features can be handled in the
Batalin-Vilkovisky  quantization procedure. However, the fact that the system
is
infinitely reducible means that an infinite tower of ghost fields are needed.
This
leads to  some problems, such as the problem of regularizing infinite sums
and the
problem of Stueckelberg type residual gauge symmetries of the final BRST
invariant
action. A period of intense activity in these area occurred in 1989,
resulting in a
number of papers to which we refer the reader for a detailed discussion of
these
problems [6][20].

	More recently, the idea of trading the $\kappa$--symetry for
world-line local
supersymmetry has been put forward [21] in attempt not only to understand the
origin
of $\kappa$--symmetry as a special supersymmetry transformation, but also to
achieve the covariant quantization in a way which may avoid the problems
mentioned
briefly above, as well as achieving an off-shell formulation which would then
enable one to construct higher order invariants. (Some progress has already
been
made in [22] towards the quantization of these theories). The approach of ref.
[21], which is currently gaining popularity, makes use of twistor-like
variables,
which are essentially commuting fermionic variables arising as superpartners
of the
fermionic coordinates of the target superspace. We now turn to a brief
discussion
of this approach.  \bigskip \centerline{\it  Twistor-like Formulation}
\bigskip
In order to introduce the twistor-like formulations, it is convenient to
pass to the first order form of the action, which for the massless
superparticle
in flat target superspace is given by
$$
S=\int d\tau\bigl[P_a\bigl(\partial_\tau X^a-
{\bar\theta}\Gamma^a\partial_\tau\theta\bigr)-\ft12 eP_aP^a\ . \eqno(2.22)
$$
This action is invariant under the $\kappa$--symmetry transformations
$$
\delta X^a ={\bar\theta}\Gamma^a\delta\theta\ ,\qquad
\delta\theta=\Gamma^aP_a\kappa\ ,\qquad
\delta e=-4{\bar\kappa}\partial_\tau\theta\ ,\qquad \delta P_a=0\ . \eqno(2.24)
$$
The constraints which follow from the action (2.22) are the reparametrization
constraint $T := P_aP^a\approx 0$, and the fermionic constraint $d_\alpha :=
P_\alpha-P_a\Gamma^a_{\alpha\beta}\theta^\beta\approx 0$, where $P_\alpha$
is the
conjugate momentum associated with $\theta_\alpha$ (see eq. (2.10)). The
fermionic
constraint is a combination of first class constraints generating the
$\kappa$--symmetry transformations and second class constraints which, of
course,
have no such an interpretation.

	The main idea of the twistor-like formulation is to replace the
momentum variable
$P_m$ with a suitable combination of {\it commuting} fermions, which are the
twistor-like variables, to reformulate the action (2.3) in a world-line locally
supersymmetric fashion [21]. The word {\it twistor-like} is used to avoid
confusion
with the supertwistor which consists of a multiplet of fields forming a
multiplet of
superconformal groups which are known to exists in dimensions $d\le 6$. In fact
such variables have been used previously in a twistor formulation of
superparticles and superstrings in $d=3,4,6$ [23]. A similar, but not quite the
same, multiplet of variables were introduced in [24] to give a twistor-like
formulation of these models in $d=10$ as well. However, the twistor-like
formulation we shall briefly review below, which is due to [21], differs
from this
formulation, in that it involves a different set of variables yet. The
advantage of
this formulation is that it allows a natural generalization to curved
superspace, as well as higher super p-branes [25][26]. It should be noted
that in
all these three different formulations, there is a common variable, namely the
commuting spinor mentioned earlier which is used to replace the momentum
variable $P_a$.

For superparticles and superstrings the twistorlike formulation works in
$d=3,4,6,10$. To simplify matters, let us focus our attention to the $d=3$
case.
The twistor-like formulation of the massless superparticle in $d=3$ is given by
[21]
$$
S=\int d\tau P_a
\bigl(\partial_\tau X^a-{\bar\theta}\Gamma^a\partial_\tau\theta +
{\bar\lambda}\Gamma^a\lambda\bigr)\ , \eqno(2.24)
$$
where $\lambda$ is the twistor-like variable, a two component commuting
Majorana
spinor in $d=3$. Its equation of motion is $P_a\Gamma^a\lambda =0$, whose
solution
can be shown to be
$$
P_a=e^{-1}{\bar\lambda}\Gamma_a\lambda\ ,  \eqno(2.25)
$$
thanks to the identity
$$
\Gamma^a_{(\alpha\beta}\Gamma^a_{\gamma)\delta}=0\ .\eqno(2.26)
$$
Note that $P_a P^a=0$ as well, due to this identity. In this action
$\kappa$--symmetry has been replaced by $n=1$ local world-line supersymmetry
generated by $Q := \lambda^\alpha d_\alpha$ [21], were we recall that
$d_\alpha=P_\alpha-\bigl(P_a\Gamma^a\theta\bigr)_\alpha$.

	To close the $n=1$ supersymmetry off-shell, it is convenient to
pass to a
superfield formalism which will naturally introduce the necessary auxiliary
fields. To this end, let us consider the superline with coordinates
$(\tau,\eta)$
and the following superfields
$$
P_a(\tau,\eta)=P_a+\eta \rho_a\ ,\qquad X^a(\tau,\eta)=X^a+\eta\chi^a\ ,
\qquad \theta(\tau,\eta) =\theta+\eta\lambda\ , \eqno(2.27)
$$
where $\eta$ is the single anticommuting coordinate, and the auxiliary fields
$\rho$ and $\chi$ have been introduced. Note that the twistor variable
$\lambda$
has been paired with the target space fermionic variable $\theta$. In terms of
these {\it superfields} the off-shell supersymmetric action is [21]
$$
S=\int d\tau d\eta P_a\bigl(DX^a+{\bar\theta} \Gamma^a D\theta\bigr)\ ,
\eqno(2.28)
$$
where $D= {\partial\over\partial\eta}+\eta{\partial\over\partial\tau}$. Note
that
this action has the form of a Wess-Zumino term.

	To formulate the massless superparticle action in $d >3$, and to put
the above
formulation in a somewhat more geometrical form that will enable us to make the
curved superspace and higher p-brane generalizations, it is useful to introduce
the following notation. For definiteness let us focus on $d=10$, in which case
the $\kappa$ symmetry can be replaced by $n=8$ world-line supersymmetry. The
coordinates of $n=8$ worldline superspace will be denoted by $Z^M=(\tau,\mu)$
and the coordinates of the target superspace by $Z^{\ul M}=({\ul m},{\ul
\mu})$,
where $\mu=1,...8,\ {\ul m}=0,1,...,9,\ {\ul \alpha}=1,...,16$. We take the
$d=10$
spinors to be Majorana-Weyl. The world-line supervielbein will be denoted by
$E_M{}^A$ and the target space supervielbein by  $E_{\ul M}{}^{\ul A}$,
where the
tangent space indices split as $A=(0,r)$ and ${\ul A}=({\ul a},{\ul\alpha})$,
respectively, with $r=1,...8,\ {\ul a}=0,1,...9,\ {\ul\alpha =1,...,16}$. Note
that the un-underlined indices always refer to the world-line superspace
quantities, while their underlined versions always refer to the corresponding
target space quantities.

  The twistor-like formulation of massless superparticle and superstrings
in $d>3$
with $\kappa$ symmetry completely replaced by an $n$--extended worldline
supersymmetry requires the introduction of $n$ twistor-like variables
$\lambda_r{}^{\ul\alpha}$ satisfying the constraint [27]
$$
{\bar\lambda}_r\Gamma^{\ul a}\lambda_s=\ft18\delta_{rs}\bigl(
            {\bar\lambda}_q\Gamma^{\ul a}\lambda_q\bigr)\ . \eqno(2.29)
$$
To express this constraint in a geometrical form, it is convenient to introduce
the notation
$$
E_A{}^{\ul A} = E_A{}^M\bigl(\partial_M Z^{\ul M}\bigr)E_{\ul M}{}^{\ul A}\ .
   \eqno(2.30)
$$
Let us furthermore make the identifications
$$
E_r{}^{\ul\alpha}\vert_{\theta=0} =\lambda_r{}^{\ul\alpha}\ ,\quad\quad\quad
E_0{}^{\ul a}\vert_{\theta=0}={\cal E}_0{}^{\ul a}\ ,\eqno(2.31)
$$
where ${\cal E}_0{}^{\ul a}=\partial_\tau X^{\ul a}-{\bar\theta}
\Gamma^{\ul a}\partial_\tau\theta $, in flat target superspace. Thus, the
fermionic
coordinates $\theta^{\ul\alpha}$ have been elevated to a superfield whose
expansion is of the form
$\theta^{\ul\alpha}(\tau,\theta)=\theta^{\ul\alpha}(\tau)
+\lambda_r{}^{\ul\alpha}(\tau)\theta^r+\cdots$. Furthermore, for
definiteness and
to simplify matters we shall characterize, both, the worldline and target
superspace geometries. Namely, we shall take the worldline superspace to be
characterized by the torsion constraints
$$
T_{rs}{}^0=2\delta_{rs}\ ,\quad\quad T_{0r}{}^0=0\ ,\quad\quad
T_{s0}{}^r=0\ ,\quad\quad T_{rs}{}^q=0\ ,  \eqno(2.32)
$$
and we shall take the target superspace geometry to be characterized by
the torsion constraints
$$
T_{\ul{\alpha\beta}}{}^{{\ul c}}=
2\gamma_{\ul{\alpha\beta}}^{\ul c}\ ,\quad\quad\quad
T_{\ul{b\alpha}}{}^{\ul a}=0\ , \quad\quad\quad
T_{\ul{\alpha\beta}}{}^{\ul\gamma}=0\ . \eqno(2.33)
$$
 Of course, all consequences of these constraints which follow from the Bianchi
identities are understood to hold.

	The $n=8$ locally supersymmetric action can now be written as [28]
$$
S=\int d\tau d^8\theta P_{\ul a}{}^r E_r{}^{\ul a}\ ,\eqno(2.34)
$$
where $P_{\ul a}{}^r$ is a Lagrange multiplier superfield. The constraints
(2.32)
leave enough room for worldline diffeomorphisms and local $n=8$ supersymmetry.
(See ref. [28] for a detailed description of these transformations).

The  field for the Lagrange multiplier is  $E_r{}^{\ul a}=0$. Taking the
curl of this equation, and recalling (2.31), yields an integrability condition
whose
$\theta=0$ component is the twistor constraint (2.29). The classical
equivalence of
this theory to the usual $\kappa$--symmetric one has been shown in [28].

Analogous twistor-like formulations have also been proposed for massive
superparticle [29][26], superstrings [28][30], supermembranes [25] and all
higher super p-branes [26]. The questions of covariant quantization, and higher
order invariants have been so far addressed only in limited cases. It has been
shown in [22] that the quantization of the superstring in which two of the
eight
$\kappa$--symmetries are replaced by $n=2$ world-sheet local supersymmetry, a
semi-light cone gauge, while not covariant, avoids a  problem  encountered in
the
usual light-cone gauge formulation due to the global issues that arise in
choosing
such a gauge. As for the construction of higher order $\kappa$--invariants,
results
have been obtained for the case of massive superparticle in $d=2,3$ [29] by
using
the twistor-like formulation. (See the references in [29] for some other
approaches to $\kappa$--symmetry calculus). We next turn to the description
of the
$\kappa$--symmetric superstring theory in curved superspace.
\bigskip
\noindent{\bf 3. Superstring in Curved Superspace}
\bigskip

For definiteness, let us focus our attention on the heterotic string
propagating in $d=10$ supergravity plus $SO(32)$ Yang-Mills background. Let the
coordinates of the world-sheet be $\sigma^i, i=0,1$, and the pulled-back
supervielbein
$$
E_i^A= \partial_i Z^M E_M^{~A}\ , \eqno(3.1)
$$
where the tangent space index splits as $A=(a,\alpha)$ with $a=0,1,...,9$ and
$\alpha=1,...,16$. We shall be dealing with sixteen component
{\it Majorana-Weyl}
spinors in $d=10$. The ingredients for the action are the Kalp-Ramond
super 2-form $B=dZ^M\wedge dZ^N B_{NM}$, the Yang-Mills superfield $A_M$, the
dilaton superfield $\phi$ and the
{\it heterotic fermions}\ \ $\psi^I, I=1,...,32$
which are {\it Majorana-Weyl} spinors on the world-sheet. In terms of these
building
blocks, the heterotic string action can be written as
$$
S=\int d^2\sigma \Bigl[-\ft12 \phi \sqrt {-g} g^{ij}E_i^a E_j^a +
\ft12 \epsilon^{ij}\partial_i Z^M\partial_j Z^N B_{NM}
+\ft12 \sqrt {-g}\psi^I\gamma^i D_i\psi^I\Bigr]\ ,\eqno(3.2)
$$
where $g_{ij}$ is the world-sheet metric of signature $(-1,+1)$, $ g={\rm det}\
g_{ij}$ and  $D_i\psi^I=\bigl(\partial_i\delta^{IJ}+
\partial_i Z^M A_M^{IJ}\bigr)
\psi^J$.  We shall require that the action be invariant under the
$\kappa$--symmetry
transformations of the form
$$
\delta Z^M=\kappa_{i\alpha}P_-^{ij}
E_j^a\Gamma_a^{\alpha\beta}E_\beta{}^M \ ,\quad\quad
\delta \psi^I=-\delta Z^M  A_M^{IJ}\psi^J\ ,\quad\quad
e_{ir}\delta e^{jr}=P_{+im}P_+^{jn}S^{m\alpha }\kappa_{n\alpha}\ ,\eqno(3.3)
$$
where $\kappa_{i\alpha}$ is the transformation parameter. Note
that unlike in the particle case, this parameter is a world-sheet vector in
addition to being a target space (Majorana-Weyl) spinor. Note also that only
its
self-dual projection  occurs. The quantity $S^{i\alpha}$ is to be determined
by the
$\kappa$--symmetry of the action. The duality projectors are defined as
$P_{\pm ij}=\ft12 \bigl(g_{ij}\pm \sqrt {-g}\epsilon_{ij}\bigr)$, satisfying
the
relation $P_+^{ki}P_{+kj}=0$.  Note that the Levi-Civita
symbols $\epsilon_{ij}$ and $\epsilon^{ij}$ are both constants, satisfying
$\epsilon^{ij}\epsilon_{jk}=\delta^i_k$.

The $\kappa$--symmetry of the action
imposes the constraints
\footnote{$^\dagger$}{\tenfoot  The study of $\kappa$--symmetric string
theory in
{\it curved} background was pioneered by Witten [8]. Soon afterwards, various
extensions of his results were obtained in refs. [31][32].}
$$
\eqalignno{
&T_{\alpha\beta}{}^c =2(\Gamma^c)_{\alpha\beta}\ ,\qquad
T_{\alpha (bc)} =u^\beta{}_{(b}\Gamma_{c)\beta\alpha}+\eta_{bc} v_\alpha\ ,
&(3.4a)\cr
&H_{\alpha\beta\gamma} =0\ ,\quad\quad
H_{a\alpha\beta}=-2\phi(\Gamma_a)_{\alpha\beta}\ , \quad  H_{ab\alpha}=
2\phi(\Gamma_{ab})_\alpha{}^\beta h_\beta
             +2\phi u^\beta{}_{[a}\Gamma_{b]\beta\alpha} \ ,&(3.4b)\cr
&F_{\alpha\beta}^{IJ}=0\ ,\qquad
F_{a\alpha}^{IJ}=(\Gamma_a)_{\alpha\beta} \chi^{\beta IJ}\  &(3.4c)\cr
&h_\alpha=v_\alpha+\ft12 \phi^{-1}D_\alpha \phi\ . &(3.4d)\cr}
$$
Furthermore, the quantity $S_i^\alpha$ is determined to be
$$
S_i^\alpha = -4E_i^\alpha +2E_i{}^a \bigl( -u^\alpha{}_a
+\Gamma_a^{\alpha\beta} h_\beta \bigr)
-\ft12 \phi^{-1}{\bar \psi}^I\gamma_i\psi^J \chi_{IJ}^\alpha  \ .\eqno(3.5)
$$
(Concerning the occurrence of $u^\alpha{}_a$ in these formulae, see the
footnote
below eq. (2.6)). The action (3.2) is also invariant under the world-sheet
reparametrization and Weyl scalings. While the rigid spacetime superconformal
symmetries are no longer possible [11], the action is of course invariant
under the
rigid spacetime Poincar\'e supersymmetry.

 Assuming the $\kappa$--symmetry constraints (3.5), the
action (3.2) is also invariant under the following local bosonic
$\lambda$--symmetry transformations
$$
\delta Z^M=P_+^{ij}E_i^\alpha \lambda_j E_\alpha{}^M\ ,\quad\quad
\delta e_{ir}=0 , \eqno(3.6)
$$
provided that $u^\alpha{}_a=0$, $h_\alpha=0$ and $\chi^\alpha_{IJ}=0$. (For the
variation of $\psi^I$ we adopt the same rule as in (2.15)). The flat
target space
version of this symmetry has been given in ref. [2].

The question
arises as to what supergravity theory, if any, do the above constraints
describe.
In the absence of the Yang-Mills sector, and ignoring the question of
$\kappa$--symmetry anomalies, it was shown by Witten that the constraints
of pure
$d=10, N=1$ supergravity are consistent with $\kappa$--symmetry. Whether
they are
necessary for $\kappa$ symmetry is more difficult to establish. To this
end, as in the particle case, one may investigate the closure of first class
constraints  which may be constructed out of the constraints which follow
from the
Lagrangian. Then allowing field redefinitions (including the scaling of the
supervielbein) and taking into account the Bianchi identities, in [12] it
has been
shown for the case of pure supergravity that Witten or Nilsson constraints,
supplemented by additional constraints needed for the description of $N=1,
d=10$
supergravity, can be derived.

However, pure $N=1, d=10$ supergravity has gravitational anomalies. One has to
include an $SO(32)$ or $E_8 \times E_8$ Yang-Mills sector, and utilize the
Green-Schwarz mechanism to cancel the gravitational, gauge and mixed anomalies.
These anomalies reflect themselves in the form of
$\kappa$--symmetry anomalies, from the world-sheet point of view, which can
also be
cancelled by a similar mechanism. In trying to understand the consequences
of the
$\kappa$--symmetry constraints (3.4), this anomaly cancellation mechanism
must be
taken into account. The $\kappa$--symmetry  anomalies have been discussed
in detail
in [33], and previously  in ref. [34]. To cancel them one adds a
counterterm $\Gamma$,  whose $\kappa$--symmetry variation is [33][34]
$$
    \delta \Gamma =\int d^2\sigma \epsilon^{ij}E_i^A E_j^B\kappa^\alpha
              {\cal K}_{\alpha BA} \ , \eqno(3.6)
$$
where ${\cal K}_{\alpha BA}$ are particular components of a super 3-form
${\cal K}$
which takes the form
$$
      {\cal K}= \omega_{3Y}-\omega_{3L}+\cdots\ , \eqno(3.7)
$$
and  $\omega_{3Y}$ and $\omega_{3L}$ are the Yang-Mills and Lorentz
Chern-Simons forms. Then the $\kappa$--variation of the classical action
plus the
anomalous variation $\delta \Gamma$  will yield the same result as in the
variation
of the classical action alone, except that the curvature $H=dB$ will now be
replaced by
$$
{\cal H}= dB+ {\cal K}\ . \eqno(3.8)
$$

If we wish to impose the so-called {\it standard} set of constraints, i.e.
those in (3.4) with  $u^\alpha{}_a$ and $v^\alpha$ set equal to zero (one could
consider any equivalent set), then the above replacement is consistent provided
that [33]
$$
(d{\cal K})_{\alpha\beta\gamma\delta}=0\ ,\qquad\qquad
(d{\cal K})_{a\alpha\beta\gamma}=0\ . \eqno(3.9)
$$
Such a ${\cal K}$ can be found, thanks to the existence of the following
relation [35]
$$
{\rm Tr} (RR)=dX+K\ , \eqno(3.10)
$$
for some 3-form $X$ and a 4-form $K$ which has the vanishing
projections $ K_{\alpha\beta\gamma\delta}=0$ and $K_{a\alpha\beta\gamma}=0$.
Since ${\rm Tr} (FF)$ has the same vanishing projections as well, it follows
that a suitable choice for ${\cal K}$ is [33]
$$
      {\cal K}= \omega_{3Y}-\omega_{3L}+X\ . \eqno(3.11)
$$
With this choice, $d{\cal K}={\rm Tr} (FF) - K $ indeed has the
vanishing projections in the $(\alpha\beta\gamma\delta)$ and
$(a\alpha\beta\gamma)$
directions as required.

In summary, the constraints needed to be imposed are the torsion constraints
(3.4) with $u^\alpha{}_a=0,\ v^\alpha=0$ and with the replacement
$H\rightarrow {\cal H}$ to be made. When the consequences of the Bianchi
identities
are taken into account as well, all the superfields get determined in
terms of the
totally antisymmetric $T_{abc}$, $\chi_\alpha$ and $\phi$. In the absence of
the
Lorentz Chern-Simons term, for example, the remaining projection of
${\cal H}$ is
found to be: ${\cal H}_{abc}=-\ft32 \phi T_{abc}+
\ft14 (\Gamma_{abc})_{\alpha\beta}{\rm Tr} (\chi^\alpha\chi^\beta)$ [16].
In this
case the resulting supergravity equations  are those of the usual $N=1, D=10$
supergravity plus Yang-Mills system. When the effect of the Lorentz
Chern-Simons
term is taken into account as outlined above, the resulting equations
describe an
anomaly free model which can accommodate stringy corrections, in particular an
$R^4$--type correction. For more details, we refer the reader to ref. [35].

So far we have considered the heterotic string in $d=10$. For a discussion of
actions for heterotic strings in $d=3,4,6$, see ref. [32], where the
occurrence of
various supergarvity/matter,  as well as {\it off-shell}
supergravity backgrounds is also discussed.  A Green-Schwarz type
action for the Type IIA superstring has been discussed in ref. [36], and
for  the
Type IIB string in ref. [31]. In the latter two cases, what has been shown
is that
the superspace constraints of on-shell Type IIA and Type IIB supergravities are
{\it sufficient} for the $\kappa$--symmetry of the string actions.
\bigskip
\noindent{\bf 4. Massive Superparticle, Supermembranes and Higher
               Super $p$--Branes}
\bigskip
Just as the action for the massless superparticle is similar to the heterotic
string action, the massive superparticle action is similar to the
action for supermembranes and higher super $p$--branes. We shall describe the
action for them in a unified manner.

A super $p$--brane moving in a superspace of $d$ bosonic dimensions will sweep
a
worldvolume of $p+1$ dimensions. Denoting the coordinates of the world-volume
by
$\sigma^i$ and the coordinates of the target superspace by $Z^M$, we again
adopt the definition (3.1), with $i=1,...,p+1$. The tangent space index again
splits as $A=(a,\alpha)$ with $a=0,1,...,d-1$ and for definiteness, we take
$\alpha$
we label the minimum dimensional spinors possible in $d$--dimensions. Thus,
it can
be said that we are considering Type 1 super $p$--branes, in the sense that
we are
considering the minimum possible target space supersymmetry. In fact,  while
the
existence of Type 2 super $p$--branes (i.e. target spaces with higher than
minimum supersymmetry) has been proposed, at present no action is known for
them,
and this is one of the most interesting open problems in the theory of super
$p$--branes.

 The action for super $p$--branes requires a Kalp-Ramond type super
$(p+1)$--form
whose pull-back is
$$
B_{i_1 \dots  i_{p+1}} = \partial _{i_1} Z^{M_1} \dots
  \partial_{i_{p+1}}  Z^{M_{p+1}} \ B_{M_{p+1} \dots  M_1}\ .\eqno(4.1)
$$
An action for super threebrane in $d=6$ flat target spacetime was first given
in
ref. [3]. Generalization to all super $p$--branes in curved target superspace
was found in ref. [4]. The action takes the form
$$
S = \int d^{p+1}\sigma
\biggl[ -\ft12 \sqrt {-g} g^{ij} \phi E_i^a E_j^a  +\epsilon^{i_1 \dots
i_{p+1}}B_{i_1 \dots  i_{p+1}}  +\ft12 (p-1)\sqrt{-g}\biggr]\ . \eqno(4.2)
$$
Note the presence of the worldvolume cosmological term for $p\ne 1$. It is
essential for the equation of motion for the metric $g_{ij}$ to yield the
nondegenerate form
$$
g_{ij}=\phi E_i^aE_j^a \ . \eqno(4.3)
$$
 Since this is an algebraic equation it can be used in establishing the
$\kappa$--
symmetry of the action. In this ``1.5'' formalism we need not vary the metric
$g_{ij}$ and the $\kappa$--transformation of the only independent variable
$Z^M$
is given by
$$
\delta Z^M=\kappa^\alpha (1+\Gamma)_\alpha{}^\beta E_\beta{}^M\ , \eqno(4.4)
$$
where the matrix $\Gamma$ is given by
$$
\Gamma = {\epsilon_p \over (p+1)!\sqrt {-g}}
\varepsilon ^{i_1\dots i_{p+1}} E^{a_1}_
  {i_1} \dots  E^{a_{p+1}}_{i_{p+1}} \Gamma _{a_1\dots a_{p+1}}\ ,\eqno(4.5)
$$
which satisfies $\Gamma^2 =1 $  provided that we use (4.3) and choose
$\epsilon_p=(-1)^{(p+2)(p+3)/4}$.
Thus, the matrix $(1+\Gamma)$ is a projection operator on-shell, and the
phenomenon of halving the fermionic degrees of freedom occurs for all super
$p$--branes.

The action (4.2) is invariant under the $\kappa$--symmetry transformation
provided
that in addition to the torsion constraints (3.4a), the following following
constraints on the super $(p+2)$--form $H=dB$ are satisfied [4]
$$
\eqalignno{
&H_{\alpha\beta\gamma A_1\cdots A_{p-1}} =0\ , &(4.6a)\cr
&H_{\alpha\beta a_p\cdots a_1}=
2\epsilon_p\phi(\Gamma_{a_1\cdots a_p})_{\alpha\beta}\ , &(4.6b)\cr
&H_{\alpha a_{p+1}\cdots a_1}=
\epsilon_p\phi(\Gamma_{a_1\cdots a_{p+1}})_\alpha{}^\beta h_\beta
+\epsilon_p\phi u^\beta{}_{[a_{p+1}}\Gamma_{a_1\cdots a_p]\beta\alpha}\ ,
&(4.6c) \cr
&h_\alpha= v_\alpha+\ft12 \phi^{-1}D_\alpha \phi\ ,&(4.6d)\cr}
$$
where $v_\alpha$ and $h^\alpha{}_a$ are arbitrary superfields. (Concerning the
occurrence of the vector-spinor $h^\alpha{}_a$ in the constraints, see the
footnote below eq. (2.6)). The Bianchi identity $dH=0$ is satisfied
provided that
the following $\Gamma$--matrix identity is satisfied:
$$
\Gamma^a_{(\alpha\beta}\Gamma^{a c_1\cdots c_p}_{\gamma\delta)}=0\ . \eqno(4.7)
$$
This identity which is crucial for the $\kappa$--symmetry of the action imposes
restrictions on possible dimensions $d$ of spacetime and on possible
values of $p$
for a $p$--dimensional extended object.  Assuming Lorentzian
signature, these restrictions are [5]:  $(p=1; d=3,4,6,10),\ (p=2;d=4,5,7,11),\
(p=3;d=6,8),\ (p=4;d=9), \ (p=5;d=10)$.  It is a very interesting fact that the
existence of the $\kappa$--symetry imposes so severe restrictions on both the
dimension of spacetime {\it and } the extension of the fundamental object. We
recall that the restriction $d\le 11$ arises for the existence of
supergravities
with a single gravitational field, and the restriction $d\le 6$ arises for the
existence of scalar supermultiplets. Since in  a gauge in which
$\kappa$--symmetry
is fixed, it is expected that a globally supersymmetric field theory in a
$(p+1)$--dimensional worldvolume emerges, the occurrence of the restrictions
$d\le
11$ and $p\le 5$ can be viewed as a consequence of an interesting fusion of
worldvolume and targetspace supersymmetries via the $\kappa$--symmetry.

Turning to the constraints (3.4a) and (4.6), only in the case of
eleven dimensional it is known rigorously that these
constraints (with the dilaton superfield redefined away) do imply the unique
$N=1,
d=11$ supergravity equations  of motion [4][37]. For the case of super
fivebrane in
ten dimensions, one expects that the corresponding equations of motion are
those of
the dual formulation of the $N=1, d=10$ pure supergravity, in which instead of
a
2-form field, a 6-form field occurs. In the remaining cases one expects that a
suitable version of supergravity theories possibly coupled to matter/Yang-Mills
supermultiplets will be described by the constraints required for the
existence of
the super $p$--brane actions. It would be of considerable interest to work out
explicitly the consequences of these constraints and to determine exactly which
supergravity theories are described by them. (See ref. [38] for a description
of
supergravity theories in diverse dimensions).  \bigskip
\centerline{\it Heterotic and Type II Super $p$--Branes}
\bigskip

  One of the most interesting open problems is to
find certain generalizations of the super $p$--brane actions. The possibility
of
super $p$--branes described by non--Poincar\'e symmetries of non-Lorentzian
target
spaces has been discussed in ref. [39]. In particular, the case of a $2+2$
worldvolume
embedded in $10+2$ dimensional target space is rather interesting, but no
action
has been found for it.  There are two other
kinds of generalizations whose existence have been established by indirect
means
and for which no actions have been written down so far. These are the
analogs of
the heterotic string, which are called the  {\it heterotic $p$--branes} (in
the sense that they describe the coupling Yang-Mills to the super $p$--brane)
[40][41], and the analogs of the Type IIA and Type IIB superstrings, which are
called
the  {\it Type II super $p$--branes} (in the sense that the target space
supersymmetry  is not the minimal $N=1$ supersymmetry, but instead an $N=2$
supersymmetry which could be of $(2,0)$ or $(1,1)$ type) [42][43].  The
existence of
these types of super $p$--branes has been deduced from the existence of certain
kinds of supersymmetric extended soliton solutions to the anomaly free $N=1$
supergravity plus Yang-Mills, Type IIA or Type IIB supergravity theories
in $d=10$. These are solutions with $(p+1)$--dimensional Poincar\'e symmetry
and
they are asymptotically flat in the internal directions. For example, the super
fivebrane solution of ref. [41] is obtained by solving the equations of
motion which
follow from the following low energy action for the bosonic degrees of
freedom of
the heterotic string in $d=10$:
$$
 S={1\over \alpha'^4}\int d^{10} x
\sqrt{-g}e^{-2\phi}\Biggl(R+4(\nabla\phi)^2-\ft13 H^2-\ft1{30}\alpha'
{\rm Tr} F^2 \Biggr)\ ,  \eqno(4.8)
$$
where the notation is self-explanatory. The solution  has $(1,0)$ type
supersymmetry on the worldvolume, and it takes the form [41]
$$
\eqalign{
&g_{ab}=\eta_{ab}\ ,\quad\quad g_{\mu\nu}=e^{2\phi}\delta_{\mu\nu}\ ,\quad\quad
e^{2\phi}=e^{2\phi_0}+8\alpha'{(x^2+2\rho^2)\over (x^2+\rho^2)^2}\ ,  \cr
&H_{\mu\nu\lambda}=-\epsilon_{\mu\nu\lambda}{}^\rho \nabla_\rho \phi\ ,
\quad\quad
A_\mu={-2\Sigma_{\mu\nu} x^\nu\over (x^2+\rho^2)}\ , \cr}\eqno(4.9)
$$
where we have used the notation of ref. [41], according to which, $a=0,1,...,5$
labels the coordinates of the fivebrane worldvolume, $\mu=5,...,9$ labels the
four Euclidean internal coordinates $x^\mu$, while $\phi_0$ is the value of the
dilaton at spatial infinity, $x^2=x^\mu x^\nu \delta_{\mu\nu}$, and $\rho$
is the
size of the instanton, and $\Sigma_{\mu\nu}$ are the
antisymmetric, self-dual 't Hooft matrices which arise in the description
of the Yang-Mills instantons in four Euclidean dimensions. The ansatz for
$A_\mu$ lies in $SU(2)$ subgroup of $E_8 \times E_8$ or $SO(32)$.

This solution
breaks the spacetime symmetries, and in particular the $d=10$ translation
group is
broken to that of the worldvolume, and {\it half} of the spacetime symmetries
are
maintained.
(There are other interesting solutions which break more than half of
the spacetime supersymmetries [44]. Presumably the usual $\kappa$--symmetric
formulations can not exist for the super $p$--branes implied by these
solutions,
but it is tempting to consider that the twistor-like formulations may
accommodate them).
The unbroken symmetries are
linearly realized, while the broken symmetries are nonlinearly realized on the
worldvolume. The massless Goldstone bosons and fermions arising in this
symmetry
breaking turn out to be the position coordinates and the heterotic
coordinates of
the fivebrane. These fields correspond to the zero modes of the wave
operator for
the fluctuations around the soliton, but they are easier to determine (at
least their number) by using index theorems. In ref. [41], it is argued
that there
are 240 fermionic zero modes (corresponding to 120 degrees of freedom) and
120 bosonic zero modes. These include the 4 translational and 8
supertranslational zero modes.

The
effective action describing the dynamics of the massless fields is presumably a
$(1,0)$ supersymmetric hyperkahler sigma model in $5+1$ dimensions. However,
its
exact form has never been spelled out as yet. Such an action would be the gauge
fixed version of an action in which the worldvolume symmetries would be
realized
covariantly, and in particular the worldvolume supersymmetry would manifest
itself
in the form of a $\kappa$--symmetry. This covariant form of the action is even
more mysterious, and it is one of the major open problems in this subject.

As for the $p$--brane solitons of the Type IIA and Type IIB supergravities, an
extensive review about them can be found in ref. [45].  An  interesting
new feature
emerging is that the zero modes implied by these solutions correspond to
vector or
antisymmetric tensor supermultiplets of $(1,1)$ or $(2,0)$  supersymmetry on
the
worldvolume, unlike in the case of Type I super $p$--branes where the zero
modes form only scalar supermultiplets [42]. The values of $p$, the
worldvolume supersymmetry $N$ and the  nature of the worldvolume supermultiplet
which arises,  corresponding to the solutions of  Type IIA supergravity
in $d=10$
are [42][45]
$$
\{p=4,N=2;\ A_\mu, 4\lambda, 5\phi\}\ ,\quad
\{p=5, N=(2,0);\ B^{-}_{\mu\nu}, 4\lambda, 5\phi \}\ ,\quad
\{p=6, N=1;\ A_\mu,\lambda, 3\phi\}\ ,
$$
while those corresponding to the solutions of Type IIB supergravity in $d=10$
[42][45] are
$$
\{p=3, N=2;\ A_\mu, 4\lambda, 5\phi\}\ ,\quad
\{p=5, N=2;\ A_\mu, 2\lambda, 4\phi\}\ ,
$$
where $\lambda$ is the fermionic partner of the worldvolume Maxwell field
$A_\mu$, and $B_{\mu\nu}^{-}$ is a worldvolume antisymmetric tensor field with
a
self-dual field strength.

There is a sense in which some of the super $p$--brane theories may be dual to
each other. A super $p$--brane in $d$ dimensions may be considered to be
dual to a
super $(d-p-3)$--brane in $d$ dimensions, when each one exists as a soliton
of  the other. In a stronger sense, dual theories describe the
strong and weak coupling regimes of the same physics, but this is rather
difficult
to prove rigorously. Among the
duality relations probably the most interesting one is the heterotic
string-heterotic fivebrane  duality [40][41][46]. The two
major open questions are how to construct the heterotic fivebrane action and
how to
covariantly quantize a $\kappa$--symmetric theory (or its twistor-like
version).

As for the actions for super $p$--branes {\it in a physical gauge},
in principle the theory of nonlinear realizations applicable to spacetime
supersymmetries can be used for their construction. This procedure has
been illustrated for some special cases involving  string solitons arising in
certain globally supersymmetric field theories [47][3]. However, application of
this
procedure to higher super $p$--branes becomes quickly rather cumbersome
[48]. Nonetheless, an interesting proposal has emerged in this area, namely,
that
the effective action for some of the super $p$--brane actions, in a certain
limit, must reduce to certain supersingleton actions which live at the
boundary of
an appropriate anti de Sitter (AdS) space [49]. (A similar proposal had
been made before according to which the action for fluctuations of a super
$p$--brane
theory compactified on a $AdS^{p+2}\times S^{d-p-2}$ were to be described
by supersingleton theories formulated at the $S^p\times S^1$ boundary of of
the AdS space [50]).

An alternative approach, which may have the additional advantage of possibly
yielding a covariant action, is based on the twistor--like formulation of the
super $p$--branes [25][26].  In this approach, the theory possesses a rigid
target
space supersymmetry and a local worldvolume supersymmetry. The
$\kappa$--symmetry
emerges as a special worldvolume supersymmetry transformation. The classical
equivalence of the twistor--like and the usual $\kappa$--symmetric
formulation of
the heterotic string has been shown to hold [28]. A similar conclusion has
been reached in ref. [25] for the case of $p=2$.
 However in [26], while we agree
with the form of the action of ref. [25] and we generalize it to all super
$p$--branes, it is by no means clear to us how this theory could be
equivalent to the usual $\kappa$--symmetric one, even for the case of $p=2$.
When
the dust settles on this issue, we will probably learn some intriguing and
highly nontrivial aspects of these theories.

There exists the
possibility that the formulation of ref. [26], after all, {\it is}
inequivalent to
the usual super $p$--brane action, and that it must be taken in its own right
as
a candidate for a novel super $p$--brane
theory. In any event, the fact that we are now dealing with worldvolume
supersymmetric field theories may give us a handle on the problem of how to
construct the so far elusive heterotic and Type II super $p$--branes actions,
as
well as the problem of how to covariantly quantize the super $p$--brane
theories.

\np
\bigskip\bigskip
\bigskip
\centerline{\bf ACKNOWLEDGEMENTS}
\bigskip
\bigskip
 I am deeply grateful to Professor Abdus Salam, in whose honor this conference
has been held, for more than a decade of continued support, encouragement and
inspiration. It is an honor and privilege to know and to have closely worked
with a great scientist, a great visionary, a great man.

I am delighted to have been involved with the activities of one of
Professor Abdus Salam's most wonderful creations, the International Centre
for Theoretical Physics in Trieste, in the capacity of a staff member for many
years. I have very fond memories of my Trieste years, working closely with
Professor Abdus Salam and benefitting greatly from his profound wisdom. I have
always been deeply impressed with his amazing ability to focus on the relevant
issues, always going straight to the heart of the matter.

Over the years, I not only  came to
appreciate  the great contributions of
Professor Abdus Salam to theoretical physics, but I also witnessed closely his
monumental contributions to the development of science in third world
countries,
through his relentless efforts, and with his continuous idea-generating
capacities.  I would like to take this opportunity to express my great
admiration
and respect for him.

\np
\centerline{\bf REFERENCES}
\bigskip\bigskip

\item{1.} W. Siegel, Phys. Lett. {\bf B128} (1983) 397.
\item{2.} M.B. Green and J.H. Schwarz, Phys. Lett. {\bf 136B} (1984) 367; Nucl.
          Phys. {\bf B243} (1984) 285.
\item{3.} J. Hughes, J. Liu and J. Polchinski, Phys. Lett. {\bf B180}
(1986) 370.
\item{4.} E. Bergshoeff, E. Sezgin and  P.K. Townsend, Phys. Lett. {\bf B189}
       (1987) 75.
\item{5.} A. Ach\'ucarro, J.M. Evans, P.K. Townsend and D.L. Wiltshire, Phys.
           Lett. {\bf B198} (1987) 441.
\item{6.} R. Kallosh, Phys. Lett. {\bf B224} (1989)273; Phys. Lett. {\bf B225}
          (1989) 49;
\item{} M. Green and C.M. Hull, Phys. Lett. {\bf 225B} (1989) 57;
\item{} S.J. Gates, M. Grisaru, U. Lindstr\"om. P. van Nieuwenhuizen, M.
          Ro\'cek, W. Siegel and A. van de Ven, Phys. Lett.
{\bf B225} (1989) 44.
\item{7.} M. Henneaux and L. Mezincescu, Phys. Lett. {\bf B152} (1985) 340.
\item{8.} E. Witten, Nucl. Phys. {\bf B266} (1986) 245.
\item{9.} Abdus Salam and J.Strathdee, Nucl. Phys.  {\bf B76} (1974) 477.
\item{10.} L. Brink and J.H. Schwarz, Phys. Lett. {\bf B100} (1981) 310.
\item{11.} E.R.C. Abraham, P.S. Howe and P.K. Townsend, Class. Quant. Grav.
                                {\bf 6}(1989) 1541.
\item{12.} J.A. Shapiro and C.C. Taylor, Phys. Rep. {\bf 191} (1990) 221.
\item{13.} W. Siegel, Nucl. Phys. {\bf B263} (1985) 285; Phys. Lett. {\bf B203}
           (1988) 79.
\item{14.} B.E.W. Nilsson, Nucl. Phys. {\bf B188} (1981) 176.
\item{15.} E. Bergshoeff, P.S. Howe, C.N. Pope, E. Sezgin and E. Sokatchev,
Nucl.
                     Phys. {\bf B354} (1991) 113.
\item{16.} J.J. Atick, A. Dhar and B. Ratra, Phys. Rev. {\bf D33} (1986) 2824.
\item{17.} G. Mack and Abdus Salam, Ann. Phys. {\bf 53} (1969) 174.
\item{18.} C. Fronsdal, Phys. Rev. {\bf D26} (1982) 1988.
\item{19.} I. Batalin and G. Vilkovisky, Phys. Rev. {\bf D28} (1983) 2567.
\item{20.} E. Bergshoeff and R.E. Kallosh, Nucl. Phys. {\bf B333} (1990) 605.
\item{21.} D.P. Sorokin, V.I. Tkach and D.V. Volkov, Mod. Phys. Lett. {\bf A4}
           (1989) 901;
\item{} D.P. Sorokin, V.I. Tkach, D.V. Volkov and A.A. Zheltukhin, Phys. Lett.
          {\bf B216} (1989) 302.
\item{22.} N. Berkovits, Nucl. Phys. {\bf B379} (1992) 96; {\bf B395} (1993)
77.
\item{23.} I. Bengston and M. Cederwall, Nucl. Phys. {\bf B302} (1988) 81.
\item{24.} N. Berkovits, Phys. Lett. {\bf B247} (1990) 45.
\item{25.} P. Pasti and M. Tonin, preprint, DFPD/93/TH/07.
\item{26.} E. Bergshoeff and E. Sezgin, preprint, CTP TAMU-28/93, UG-5/93.
\item{27.} A.S. Galperin, P.S. Howe and K.S. Stelle, Nucl. Phys. {\bf B368}
(1992)
           281.
\item{28.} F. Delduc, A. Galperin, P. Howe and E. Sokatchev, Phys. Rev.
{\bf D47}
            (1993) 578.
\item{29.} J.P. Gauntlett, Phys. Lett. {\bf B272} (1991) 25.
\item{30.} M. Tonin, Phys. Lett. {\bf B266} (1991) 312; Int. J. Mod. Phys. {\bf
A7} (1992) 6013;
\item{} N. Berkovits, Nucl. Phys. {\bf B379} (1992) 96;
\item{} D.P. Sorokin and M. Tonin, preprint, DFPD/93/TH/52.
\item{31.} M.T. Grisaru, P.S. Howe, L. Mezincescu, B.E.W. Nilsson and P.K.
                          Townsend, Phys. Lett. {\bf B162} (1985) 116.
\item{32.} E. Bergshoeff, E. Sezgin and P.K. Townsend,
           Phys. Lett. {\bf 169B} (1986) 191.
\item{33.} M. Tonin, Int. J. Mod. Phys. {\bf 3}(1988) 1519;
           Int. J. Mod. Phys. {\bf 4}(1989) 1983.
\item{34.}  J.J. Atick, A. Dhar and B. Ratra, Phys. Lett. {\bf B169} (86) 54.
\item{35.} L. Bonora, P. Pasti and M. Tonin, Phys. Lett. {\bf B188} (1987) 335.
\item{36.} S. Bellucci, S.J. Gates, Jr., B. Badak and S. Vashakidze, Mod. Phys.
           Lett. {\bf A4} (1989)1985.
\item{37.} M.J. Duff, P.S. Howe, T. Inami and K.S. Stelle, Phys. Lett.
{\bf B191}
              (1987) 70.
\item{38.} Abdus Salam and E. Sezgin, {\it Supergravities in Diverse
           Dimensions} (World Scientific, 1989).
\item{39.} M. Blencowe and M.J. Duff, Nucl. Phys. {\bf B310} (1988) 387.
\item{40.} M.J. Duff, Class. Quant. Grav. {\bf 5} (1988) 189.
\item{41.} A. Strominger, Nucl. Phys. {\bf B343} (1990) 167. [Erratum:
{\bf B353}
          (1991) 565].
\item{42.} C. Callan, J. Harvey and A. Strominger, Nucl. Phys. {\bf B359}
(1991)
           611.
\item{43.} M.J. Duff and J.X. Lu, Nucl. Phys. {\bf B390} (1993) 276.
\item{44.} J. Harvey and A. Strominger, Phys. Rev. Lett. {\bf 66} (1991) 549;
\item{} R. G\"uven, preprint, 1991.
\item{45.} M.J. Duff and J.X. Lu, preprint, CERN-TH.6675/93, CTP/TAMU-54/92.
\item{46.} M.J. Duff and J.X. Lu, Phys. Rev. Lett. {\bf 66} (1991) 1402; Nucl.
           Phys. {\bf B357} (1991) 534.
\item{47.} J. Hughes, J. Liu and J. Polchinski, Nucl. Phys. {\bf B278}
(1986) 147.
\item{48.} E.A. Ivanov and A.A. Kapustnikov, Phys. Lett. {\bf B252} (1990) 212.
\item{49.} G.W. Gibbons and P.K. Townsend, preprint, DAMTP/R-93/19.
\item{50.} M.P. Blencowe and M.J. Duff, Phys. Lett. {\bf B203} (1988) 229.
\item{} H. Nicolai, E. Sezgin and Y. Tanii, Nucl. Phys. {\bf B305} (1988) 483.

\end